# $^{133}$Cs-NMR study on aligned powder of competing spin chain compound $Cs_2Cu_2Mo_3O_{12}$

**A Yagi[1] , K Matsui[1] T Goto[1], M Hase[2] and T Sasaki[3]**

[1] Sophia University, Physics Division, Tokyo, 102-8554, Japan
[2] National Institute for Materials Science (NIMS), Tsukuba, 305-0047, Japan
[3] Institute for Material Research (IMR), Tohoku University, Sendai, 980-8577, Japan

gotoo-t@sophia.ac.jp

**Abstract**.

$S$ = 1/2 competing spin chain compound $Cs_2Cu_2Mo_3O_{12}$ has two dominant exchange interactions of the nearest neighbouring ferromagnetic $J_1$= 93 K and the second nearest neighbouring antiferromagnetic $J_2$= +33 K, and is expected to show the nematic Tomonaga-Luttinger liquid (TLL) state under high magnetic field region. The recent theoretical study by Sato *et al.* has shown that in the nematic TLL state, the spin fluctuations are expected to be highly anisotropic, that is, its transverse component is suppressed. Our previous NMR study on the present system showed that the dominant contribution to nuclear spin relaxation comes from the longitudinal component. In order to conclude that the transverse component of spin fluctuations is suppressed, the knowledge of hyperfine coupling is indispensable. This article is solely devoted to investigate the hyperfine coupling of $^{133}$Cs-NMR site to prove that the anisotropic part of hyperfine coupling, which connects the nuclear spin relaxation with the transverse spin fluctuations is considerably large to be $A_{\mathrm{an}} = +770\ \mathrm{Oe}/\mu_{\mathrm{B}}$.

## 1. Introduction

The ground state of the competing spin chain, described by the $J_1$- $J_2$ model with two dominant exchange interactions of the nearest-neighboring ferromagnetic $J_1$ and the second nearest-neighboring antiferromagnetic $J_2$ has recently attracted much interest, because the existence of nematic Tomonaga-Luttinger state has been proposed theoretically[1-7]. The detection of this state, however requires to probe the four-spin correlation functions with high accuracy[1,2], so that it has been considered to be rather difficult. We address this problem with an assist of recent theoretical proposal by Sato *et al.*[2], who have proved that the nematic and ordinary TLL shows a distinct difference in the temperature dependence of NMR-$T_1$ in low temperatures. That is, while $T_1^{-1}$ for the latter is contributed both from the longitudinal and transverse spin correlations as $T^{2K-1}$ and $T^{1/2K-1}$, respectively, where $K$ is the Luttinger parameter[8-12], $T_1^{-1}$ for the former is contributed only from the longitudinal spin correlations. This clearly states that an experimental evidence for the nematic TLL state can be obtained, if only one observes that $T_1^{-1}$ decreases with lowering temperature under moderately high magnetic field, where $K$ becomes larger than 1/2 [8-12].

We have recently reported that $T_1^{-1}$ for the model compound $Cs_2Cu_2Mo_3O_{12}$ (Fig. 1) exhibits nematic-TLL-like temperature dependence described above[13,14]. However, in order to conclude that the transverse spin fluctuations are suppressed, one must prove that there exists a finite hyperfine coupling. Generally, the nuclear spin relaxation is driven by the transverse component of field fluctuations at the nuclear site. The nuclear spin relaxation proceeds via both types of spin fluctuations, transverse and longitudinal, only when the hyperfine coupling bears a finite anisotropic part; in that case, the direction of fluctuating field changes from the electronic site to the nuclear site. In our previous NMR report[15,16], we have studied on the randomly oriented powder (Fig. 2) to obtain the hyperfine coupling constant $A_{\rm eff} = -270\,{\rm Oe}/\mu_{\rm B}$, which, however, is an averaged value for powder. For the discussion on spin nematic TLL state, it is indispensable to determine both the isotropic and anisotropic part of hyperfine coupling tensor. This article is solely devoted to its determination; we have utilized the uniaxially aligned powder sample and measured the paramagnetic NMR shift in the wide temperature region to determine separately the two components in the hyperfine coupling tensor of Cs-site.

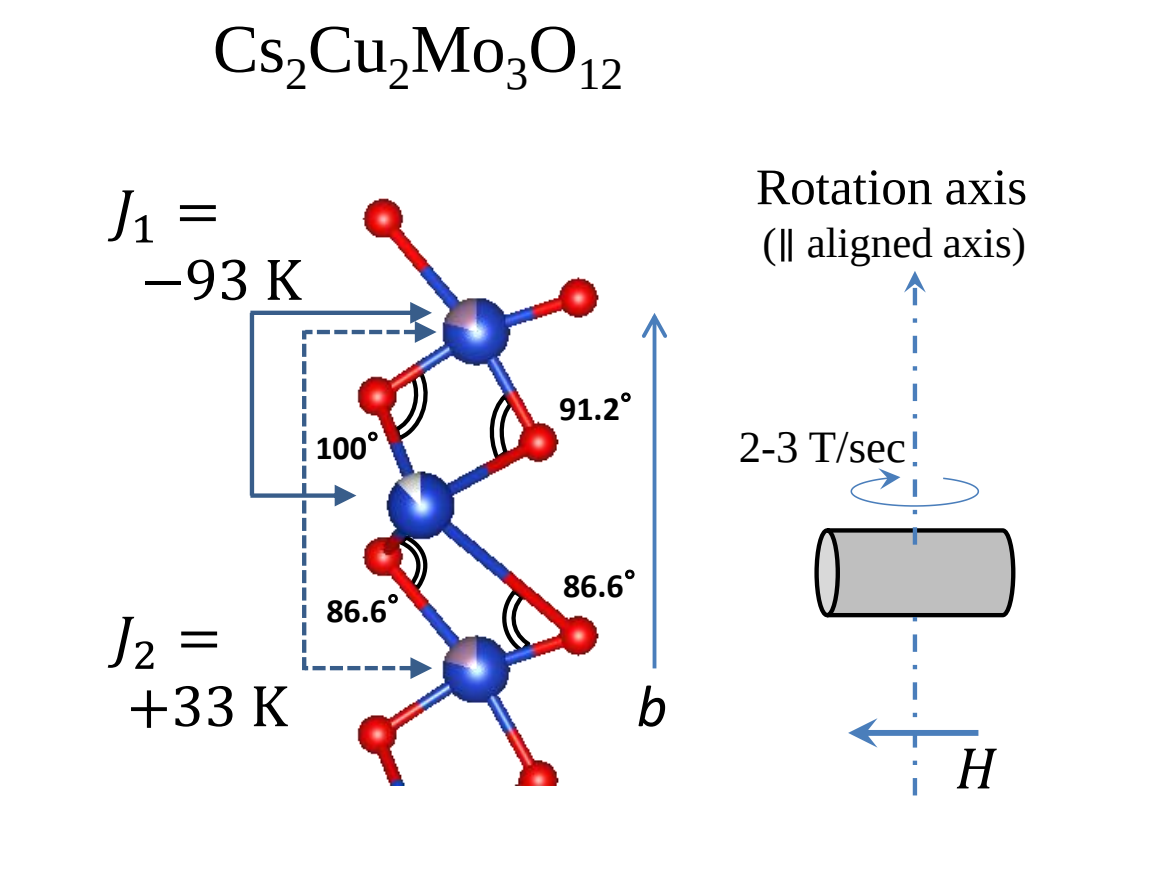

**Figure 1.** Schematic crystal structure of $Cs_2Cu_2Mo_3O_{12}$ (left) and the method of powder alignment for materials with easy plane anisotropy (right).

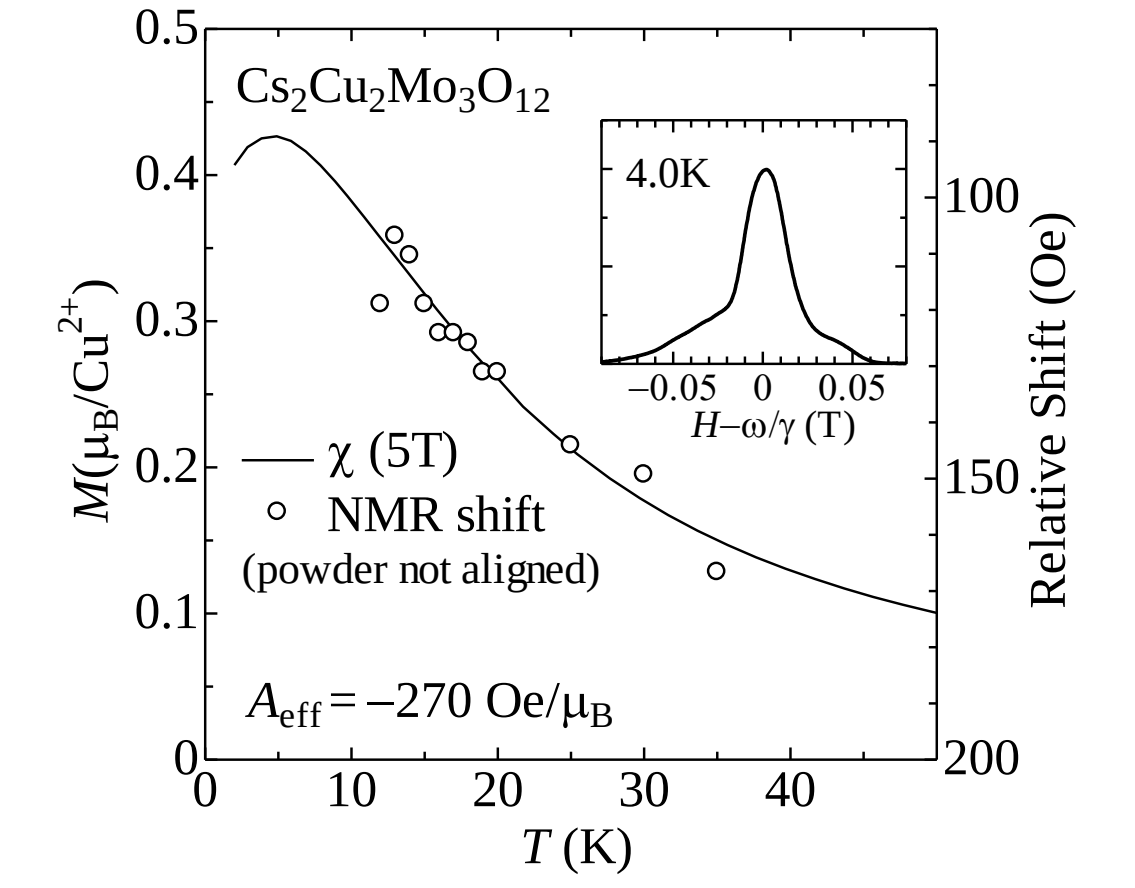

**Figure 2.** $K$-$\chi$ plot for the non-aligned powder sample[15]. The relative shift position is defined as the peak top of spectrum (inset).

Before describing the experimental method, we summarize the basic properties so far reported for the target compound $Cs_2Cu_2Mo_3O_{12}$ [13,14]. This quasi-one dimensional system is isomorphic to $Rb_2Cu_2Mo_3O_{12}$ [14], and is described by the $J_1$- $J_2$ model with $J_1$ = −93 K and $J_2$ = +33 K (Fig. 1) [13]. It is gapless under zero field, and has the saturation field $H_S$ = 9.15 T [16]. Due to a possible tiny inter-chain interaction, there is a long range magnetic order at $T_N$ =1.85 K [15-19]. The nematic TLL state is expected to exist in a finite temperature region above $T_N$.

## 2. Experimental

$Cs_2Cu_2Mo_3O_{12}$ polycrystalline sample was synthesized by the conventional solid state reaction method [13,14]. In order to determine the anisotropic NMR shift, a uniaxially-aligned powder sample was prepared by the conventional method utilizing the paramagnetic anisotropy. The powder mixed with Stycast1266, the epoxy resin, was put into a Teflon tube, and was cured in the 6T horizontal magnetic field in the room temperature, while rotating slowly (2-3 turns/sec) the tube within a plane parallel with the field (Fig. 1). By this rotational method we have successfully aligned the paramagnetic powder with the easy-plane-type anisotropy. We note here that the powder did not align well when it was cured under field without rotation.

NMR measurements for $^{133}$Cs ($I = 7/2$) spectra were performed under field region 5 - 6 T and in the temperature range between 5 and 45 K. The uniform susceptibility was measured by SQUID magnetometer (Quantum Design Co. Ltd.) under the field of 5 T.

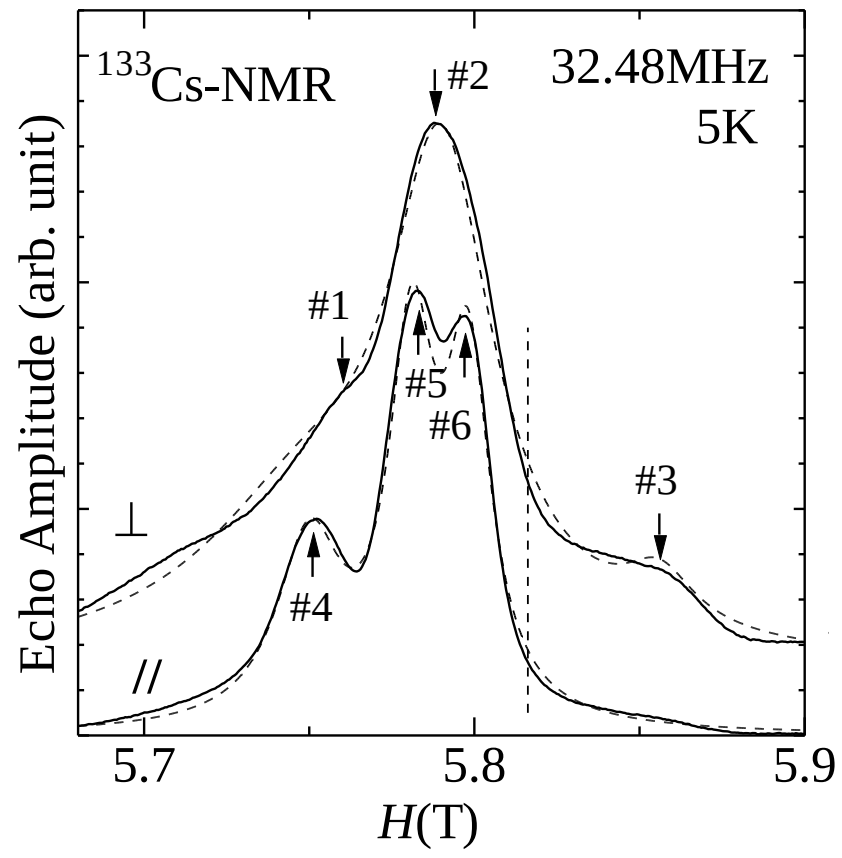

**Figure 3.** The comparison of two spectra for aligned powder sample with the field applied along or perpendicular to the aligned axis. Peak positions are indicated by arrows. The vertical line shows the zero shift position. Dashed curves show the sum of three Lorentzian peaks determined by curve fitting.

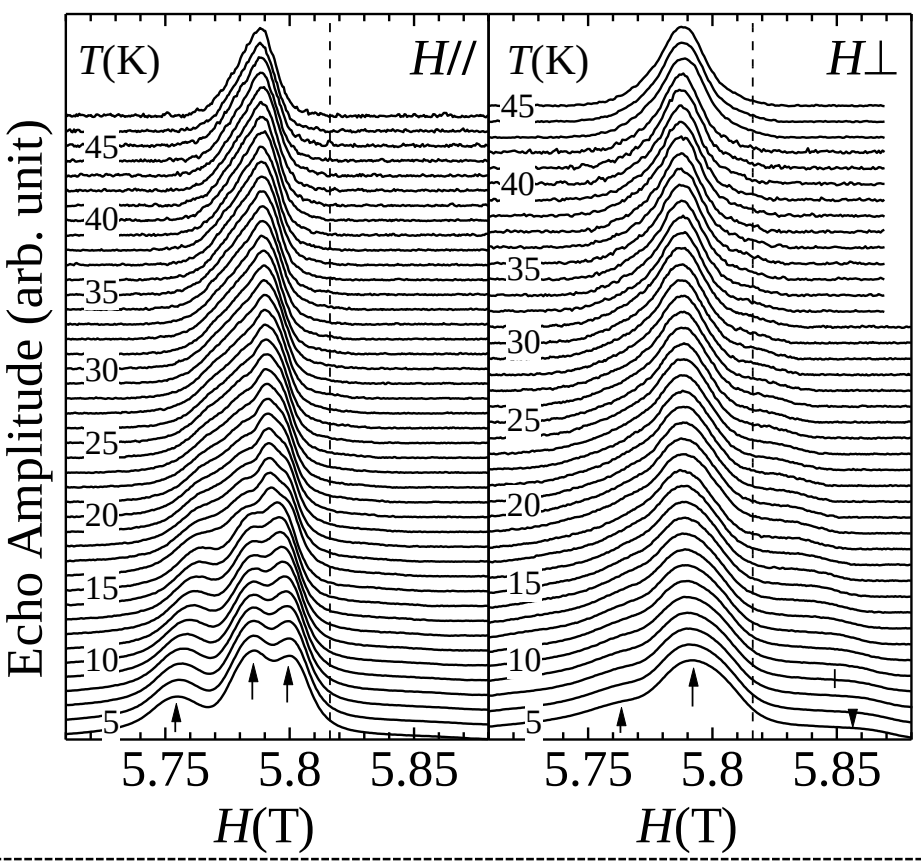

**Figure 4.** Temperature variation of the spectra of the aligned powder under field applied along parallel (*H//*) and perpendicular (*H*⊥) of the aligned axis. Arrows indicate the position of each peak.

## 3. Results and Discussion

Figure 3 shows the spectra measured under the two field directions, parallel ($H_{\parallel}$) or perpendicular ($H_{\perp}$) with the aligned axis. Note the distinct difference in the two profiles. Each two spectra in Fig. 3 are well described by a sum of the three Lorentzian peaks, denoted as #1 - #3 for $H_{\perp}$ and #4 - #6 for $H_{\parallel}$, respectively. The Lorentzian shape form comes simply from an inhomogeneity in the magnetic shift and also in the small quadrupolar splitting, which is negligibly small and hidden in each peak. These three peaks are considered to correspond to the three Cs sites in unit cell, 4*d*, 4*e* and 8*f*. The symmetric shape of each peak assures that the powders are aligned well, and also that the hyperfine tensor of each Cs site possesses a uniaxial asymmetry. We determined the peak position by curve fitting and traced the temperature dependence of peak shift in the paramagnetic region up to 45 K as shown in Fig. 4 and 5. Their temperature dependences differ from one another, including the sign of gradient. That is, with decreasing temperatures, the shifts of #1 and #4 increase, while #3 and #6 decrease. This change in the sign clearly demonstrates the existence of anisotropic part within the hyperfine coupling. We proceed to the quantitative arguments in the following.

The NMR shift with the uniaxial-anisotropy is generally described as $K_{\parallel} = K_{\rm iso} + 2K_{\rm an}$ and $K_{\perp} = K_{\rm iso} - K_{\rm an}$, where $K_{\parallel}$ and $K_{\perp}$ are the magnetic shift under the field applied parallel or perpendicular to the principle axis of anisotropy, and $K_{\rm iso}$ and $K_{\rm an}$ are, as in usual denotation, the isotropic and anisotropic part of NMR shift tensor, respectively. So, for the case where $K_{\rm iso}$ is not very large compared to $K_{\rm an}$, $K_{\parallel}$ and $K_{\perp}$ are expected to have the temperature variation with ratio of nearly −2. This is what one observes for #3 and #4, indicating that the magnetic shifts of $\Delta H_3^{\perp}$ and $\Delta H_4^{\parallel}$ correspond to $K_{\parallel}$ and $K_{\perp}$, respectively. We also note here that the aligned axis is perpendicular to the principle axis of anisotropy. We have assumed that #3 and #4 belong to the identical Cs site and tentatively calculated the $K_{\rm iso}$ and $K_{\rm an}$, which are shown in Fig. 6. By scaling them with the uniform susceptibility, we further obtained the hyperfine coupling constants of isotropic and anisotropic parts as $A_{\rm iso} = -270\,{\rm Oe}/\mu_{\rm B}$ and $A_{\rm an} = -770\,{\rm Oe}/\mu_{\rm B}$. Note that the former just coincides with that obtained for the non-aligned powder sample. And also, the smallness of $A_{\rm iso}$ compared with $A_{\rm an}$ validates our assumption above.

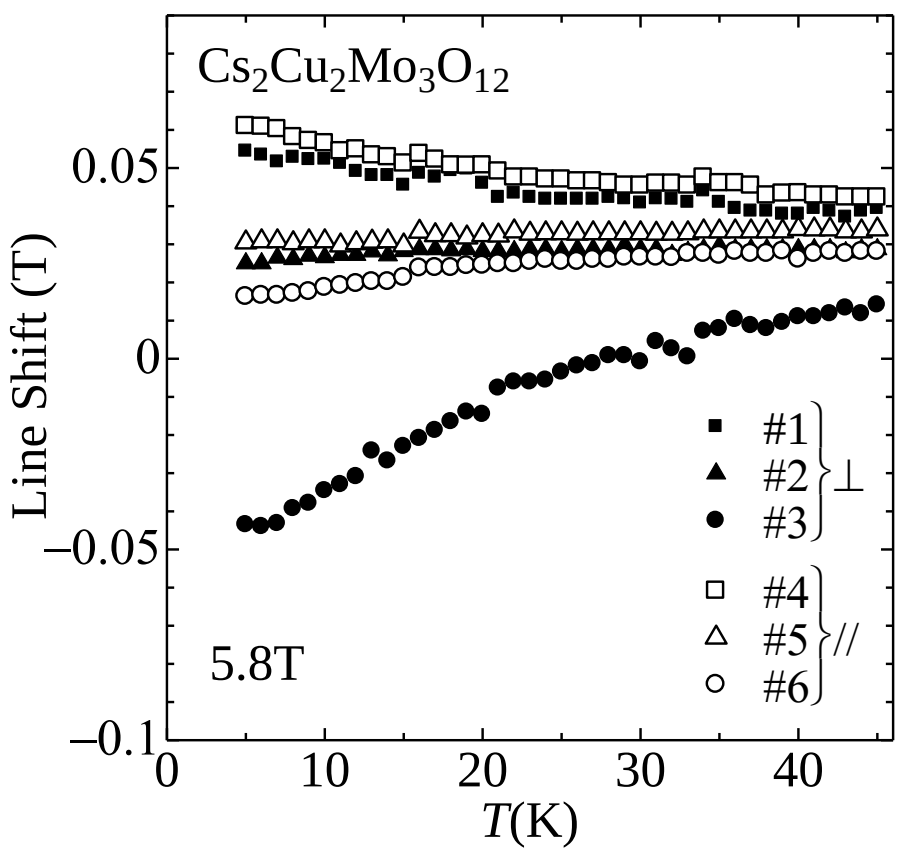

**Figure 5.** The temperature dependence of shift for each peak observed in Fig. 4.

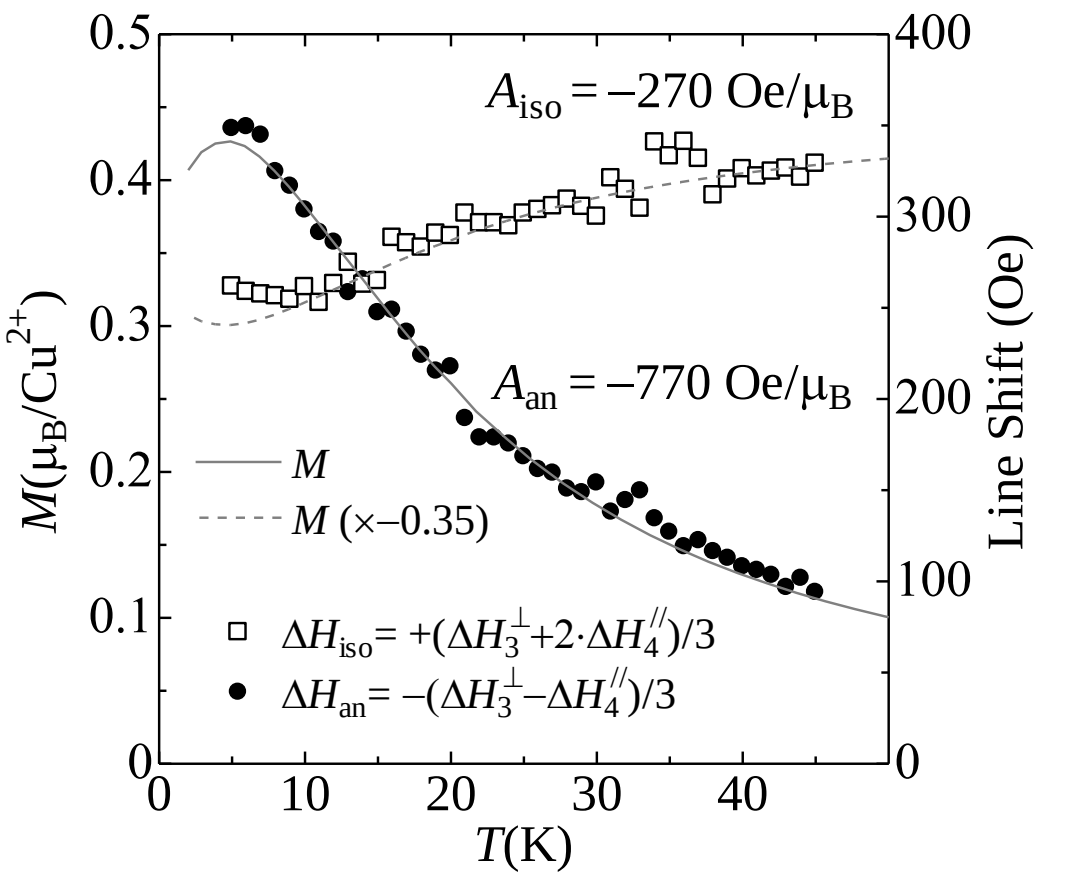

**Figure 6.** The temperature dependence of the isotropic and anisotropic shift deduced from the two-field configurations. Curves show uniform magnetization, scaled with the shifts.

Here, this considerably large $A_{an}$ indicate that the transverse component of spin fluctuations must be considerably suppressed, compared with the longitudinal one, to explain our previous observation of NMR-$T_1$, and hence, we can conclude the likeliness of the system being in the nematic TLL state[1,2,19].

Finally, we refer on the hyperfine coupling of the other two Cs-sites, and the origin of anisotropic hyperfine coupling. First, as for the other two sites, the value of hyperfine coupling constants must be much smaller than those for the Cs site assigned to #3 and #4, because they have smaller temperature dependence of the shift, as can be seen in Fig. 5. This will bring the considerable distribution in the relaxation rate among the three Cs sites. In fact, the nuclear spin relaxation for the non-aligned powder sample becomes quite inhomogeneous, and described by the stretched relaxation curve [15,17]. Next, as for the origin of $A_{an}$, a preliminary calculation has shown that the classical dipole-dipole interaction does not account for it, and that one must take into account the exchange interaction via *p* orbitals of Cs ion. Detailed calculation of dipole tensor at Cs sites for the complete site assignment including a determination of principle axes of NMR shift tensor and also measurements of $T_1^{-1}$ for each site are now in the progress.

**Acknowledgement**

This work was partly supported by JSPS KAKENHI Grant Number 15K05148.